\documentclass[preprint2]{aastex}

\shorttitle{Advection of Supergranules}
\shortauthors{Hathaway et al.}

\begin{document}


\title{The Advection of Supergranules by the Sun's Axisymmetric Flows}
\author{David H. Hathaway}
\affil{NASA Marshall Space Flight Center, Huntsville, AL 35812 USA}
\email{david.hathaway@nasa.gov}

\author{Peter E. Williams}
\affil{NASA Goddard Space Flight Center, Greenbelt, MD 20771 USA}
\email{peter.williams@nasa.gov}

\author{Kevin Dela Rosa}
\affil{School of Computer Science, Carnegie Mellon University, Pittsburgh, PA 15213 USA}

\author{Manfred Cuntz}
\affil{Department of Physics, University of Texas at Arlington, Arlington, TX 76019 USA}
\email{cuntz@uta.edu}

\begin{abstract}
We show that the motions of supergranules are consistent with a model in which
they are simply advected by the axisymmetric flows in the Sun's surface shear layer.
We produce a 10-day series of simulated Doppler images at a 15-minute cadence that
reproduces most spatial and temporal characteristics seen in the SOHO/MDI Doppler data.
Our simulated data have a spectrum of cellular flows with just two components --
a granule component that peaks at spherical wavenumbers of about 4000 and a
supergranule component that peaks at wavenumbers of about 110.
We include the advection of these cellular
components by the axisymmetric flows -- differential rotation and meridional flow -- whose
variations with latitude and depth (wavenumber) are consistent with observations.
We mimic the evolution of the cellular pattern by introducing random variations to
the phases of the spectral components at rates that reproduce the levels
of cross-correlation as functions of time and latitude. Our simulated data do not
include any wave-like characteristics for the supergranules yet can reproduce
the rotation characteristics previously attributed to wave-like behavior.
We find rotation rates which appear faster than the actual rotation rates and
attribute this to the projection effects.
We find that the measured meridional flow does accurately represent
the actual flow and that the observations indicate poleward flow to $65\degr-70\degr$
latitude with equatorward counter cells in the polar regions.
\bigskip
\end{abstract}


\section{Introduction}

Supergranules are cellular flow structures observed in the solar photosphere with typical
diameters of about 30 Mm, lifetimes of about one day, and flow velocities of ~300 
$\rm m\ s^{-1}$ \citep{RieutordRincon10}.
They cover the entire surface of the Sun except for the immediate surroundings of sunspots.

Supergranules were discovered by \citet{Hart54} but it was \citet{Leighton_etal62}
who suggested that these cellular flows were convective structures and coined
the term ``supergranule.'' However, \citet{Worden75}, and others since, had difficulty
detecting any associated thermal features (i.e. hot cell centers) consistent with a
thermal convection origin. Recently however, \citet{Meunier_etal07} have reported the
detection of a small temperature excess at cell centers of about 0.8-2.8 K.

Supergranules are intimately involved with the structure and evolution of the magnetic
field in the photosphere. \citet{SimonLeighton64} found that the magnetic structures of the chromospheric network are located at the boundaries of these cells.
\citet{Leighton64} quickly suggested that the random-walk of magnetic elements by the
evolving supergranules could transport following polarity elements poleward to reverse
the Sun's polar field each solar cycle while leading polarity elements would be
transported across the equator where they would meet and cancel their opposite
polarity counterparts.
While we now recognize that Leighton over estimated the effective diffusivity implied
by this random walk (and that the Sun's poleward meridional flow provides the missing
transport), supergranule diffusion is still not fully constrained or understood.
Leighton's initial estimate of $770-1540\ \rm km^2 \ s^{-1}$ for the diffusivity
was reduced to $200-400\ \rm km^2 \ s^{-1}$ by \citet{Mosher77} who studied the
displacements of chromospheric network elements.
\citet{SchrijverMartin90} measured the displacements of the magnetic
elements themselves using magnetograms
and found a value of  $250\ \rm km^2 \ s^{-1}$ in plage surrounding
active regions but only  $110\ \rm km^2 \ s^{-1}$ in the active regions themselves.
Models for the surface magnetic flux transport typically use values about twice this size --
\citet{vanBallegooijen_etal98} used $450\ \rm km^2 \ s^{-1}$ while \citet{Wang_etal02} used
$500\ \rm km^2 \ s^{-1}$.

Supergranules are embedded in a surface shear layer of their own making.
\citet{FoukalJokipii75} and \citet{Foukal79} suggested that the radial flows within
the supergranules would conserve angular momentum and produce a shear layer with
slower rotation at the surface and more rapid rotation below -- consistent
with the earlier observations of \citet{HowardHarvey70} that the sunspots
rotate about 5\% faster than the surrounding surface plasma.
The production of this shear layer was succesfully modelled in spherical shells by
\citet{GilmanFoukal79} and in plane-parallel layers by \citet{Hathaway82}.
The theoretical arguements suggest a change in rotation frequency such that
$\delta\Omega \sim -2\Omega \delta r/r$. In addition,
\citet{Hathaway82} found that a meridional flow was also produced with poleward
flow near the surface and equatorward flow at the bottom.

The surface shear layer has been probed with helioseismic techniques which
confirm the existence of the layer but give somewhat inconsistent results for the
detailed structure in depth and latitude.
\citet{Thompson_etal96} used the global oscillation modes from GONG and found that the
rotation rate increased inward to a depth of about 35 Mm ($\delta r/r \sim 5\%$)
and that the rate of increase was larger at the equator than at $30\degr$ latitude.
\citet{Schou_etal98} found similar results from global modes in MDI data but
noted that the shear appeared to disappear by $60\degr$ latitude and perhaps
reverse sign at higher latitudes.

\citet{Basu_etal99} used the local
helioseismology technique of ring-diagram analysis. They found a thin (4 Mm)
outer shear layer in which the shear did not appear to reverse at high latitudes and
a deeper shear layer in which it did. They also found a poleward meridional flow
which increased in amplitude across the thin outer shear layer but then
remained nearly constant across the inner shear layer. Their meridional flow
had peak velocities of about $50 \ \rm m\ s^{-1}$ at $50\degr$ latitude.
\citet{Giles_etal97} used a time-distance local helioseismology technique and
found a meridional flow with a peak velocity of only about $24 \ \rm m\ s^{-1}$ at
$45\degr$ latitude and indicated that it was constant to a depth of about 26 Mm.

The observed rotation of the supergranules has added further mystery to their nature.
(The mechanism that selects the characteristic size of supergranules is
perhaps the biggest mystery.)
\citet{Duvall80} cross-correlated the equatorial Doppler velocity patterns
and found that the supergranules rotated more rapidly than the plasma
at the photosphere and that even faster rotation rates were obtained when there
were longer (24-hour vs. 8-hour) time intervals between Doppler images.
He attributed this behavior to the surface shear layer in which larger,
longer-lived, cells extend deeper into the more rapidly rotating layers.
\citet{SnodgrassUlrich90} used
data from Mount Wilson Observatory to find the rotation rate at different
latitudes and noted that the rotation rates for the Doppler pattern were some 4\%
faster than the spectroscopic rate and, mysteriously, some 2\% faster than the
small magnetic features that are observed to outline their borders.
More recently \citet{BeckSchou00} used a Fourier transform method to find that
the larger features do rotate more rapidly than the smaller features and that the low
wavenumber components of the Doppler pattern rotate more rapidly than the plasma at
any level within the surface shear layer. This led them to
suggest that supergranules have wave-like characteristics with a preference
for prograde propagation.

\citet{Hathaway_etal06} showed that this ``super-rotation'' of
the Doppler pattern could be attributed to projection effects associated with the
Doppler signal itself. As the velocity pattern rotates across the field of view the observed
line-of-sight component is modulated in a way that essentially adds another half wave to
the pattern and gives a higher rotation rate that increases proportionally with
decreasing wavenumber.
They took a fixed velocity pattern (which had spatial characteristics that matched the
SOHO/MDI data) and rotated it rigidly to show this ``super-rotation'' effect.
While this indicated that the Doppler projection effect should be accounted for,
the fixed pattern could not account for all the variations reported by
\citet{BeckSchou00}.

\citet{Gizon_etal03} used time-distance helioseismology to
find the supergranular flow (rather than direct Doppler measurements) and
\citet{Schou03} ``divided-out'' the line-of-sight modulation. Both studies found slower
rotation rates that matched that of the magnetic features but saw evidence for
wave-like prograde and retrograde moving components.
\citet{MeunierRoudier07} compared rotation rates obtained by tracking Doppler
features, magnetic features, and divergence features (which were, in turn, derived from
correlation tracking of smaller intensity features). While they concluded that projection
effects do influence the rotation rate determined from the Doppler features, they
found that the magnetic features rotate more slowly than the supergranules.

In this paper we report on our analyses of data from the SOHO/MDI instrument
[\citet{Scherrer_etal95}] and from
simulated data in which the supergranules are simply advected by differential rotation
and meridional flow that vary with latitude and depth.
The simulated data are designed to faithfully mimic the SOHO/MDI data [the same data
that was analyzed in \citet{BeckSchou00} and \citet{Schou03}] with simple assumptions
about the dynamical structure of the surface shear layer.
The analyses include reproductions of those done in earlier studies.
Through the simulations we can better determine the actual
differential rotation and meridional flow profiles consistent with the Doppler
observations (which are subject to line-of-sight projection effects). 

\section{The Data}

The full-disk Doppler images from SOHO/MDI are obtained at a 1-minute
cadence to resolve the temporal variations associated with the p-mode oscillations.
We [cf. \citet{Hathaway_etal00} and \citet{BeckSchou00}] have temporally filtered the
images to remove the p-mode signal by using a 31-minute long tapered Gaussian with a FWHM
of 16 minutes on sets of 31 images that were de-rotated to register each to the central image.
These filtered images were formed at 15-minute intervals over the 60-day
MDI Dynamics Run in 1996. This filtering process effectively removes the p-mode
signal and leaves behind the Doppler signal from flows with temporal variations
longer than about 16 minutes.

Supergranules, with typical wavenumbers of about 110, are very well
resolved in this data (at disk center wavenumbers up to 1500 are resolved). While
granules are not well resolved, they do appear in the data as pixel-to-pixel and
image-to-image ``noise,'' as a convective blue shift (due to the correlation between
brightness and updrafts), and as resolved structures for the largest members.
These data are prepared for studying the cellular features by first measuring
and removing the Doppler signals due to: 1) observer motion, 2) convective blue
shift, 3) differential rotation, and 4) the axisymmetricmeridional flow.
The data are then mapped
onto heliographic coordinates with equal spacing in both longitude and latitude.
This mapping includes accounting for the position angle and tilt angle of the
Sun's rotation axis as well as the orientation of the SOHO/MDI detector. Note that
the position angle and tilt angle of the Sun's rotation axis determined
by \citet{Carrington63} have been found to be in error by about $0.1\degr$ by
\citet{BeckGiles05} and by \citet{HathawayRightmire10}. This correction is included
in this study.

The simulated data are constructed in the manner described previously
\citep{Hathaway88, Hathaway92, Hathaway_etal00, Hathaway_etal02}
from vector velocities with

\begin{equation}
V_r(\theta,\phi) = \sum_{l=0}^{lmax} \sum_{m=-l}^l R_l^m Y_l^m(\theta,\phi)
\end{equation}

\begin{eqnarray}
V_\theta(\theta,\phi) & = & \sum_{l=1}^{lmax} \sum_{m=-l}^l 
S_l^m {\partial Y_l^m(\theta,\phi) \over \partial \theta} + \nonumber \\
& & \sum_{l=1}^{lmax} \sum_{m=-l}^l
T_l^m {1 \over \sin\theta} {\partial Y_l^m(\theta,\phi) \over \partial \phi}
\end{eqnarray}

\begin{eqnarray}
V_\phi(\theta,\phi) & = & \sum_{l=1}^{lmax} \sum_{m=-l}^l
S_l^m {1 \over \sin\theta} {\partial Y_l^m(\theta,\phi) \over \partial \phi} - \nonumber \\
& & \sum_{l=1}^{lmax} \sum_{m=-l}^l
T_l^m {\partial Y_l^m(\theta,\phi) \over \partial \theta}
\end{eqnarray}

\noindent where $Y_\ell^m(\theta,\phi)$ is the spherical harmonic function of
degree $\ell$ and azimuthal order $m$,             
$\theta$ is the colatitude measured southward from the north pole, and
$\phi$ is the longitude measured prograde from the central meridian or some fixed
longitude. The complex quantities $R_l^m$, $S_l^m$, and $T_l^m$ are the spectral
coefficients for the radial, poloidal, and toroidal components, respectively.

To simulate the observed line-of-sight velocity, the spectral coefficients
from $\ell=0$ to $\ell=1500$ are specified and the three vector velocity
components are calculated using Eqns. 1-3
on a grid with 1500 points in latitude and 4096 points in longitude.
The Doppler velocity image is constructed by determining the longitude and
latitude at a point on the image, finding the vector velocity at that point
using bi-cubic interpolation, and then projecting that vector velocity
onto the line-of-sight using

\begin{eqnarray}
V_{los}(\theta,\phi)  & = & V_r(\theta,\phi) \sin B_0 \ \cos \theta \nonumber + \\ 
& & V_r(\theta,\phi) \cos B_0 \ \sin \theta \ \cos \phi + \nonumber \\
& & V_\theta(\theta,\phi) \sin B_0 \ \sin \theta - \nonumber \\
& & V_\theta(\theta,\phi) \cos B_0 \ \cos \theta \ \cos \phi + \nonumber \\
& & V_\phi(\theta,\phi) \cos B_0 \ \sin \phi
\end{eqnarray}

\noindent where $B_0$ is the latitude at disk center (or equivalently the tilt
of the Sun's north pole toward the observer) and velocities away from
the observer are taken to be positive. The line-of-sight velocity at
an array of 49 points within each pixel is determined and an average
is taken to simulate the integration over a pixel with the MDI
instrument.

With the current simulations the instrumental blurring is treated in a more
realistic manner. Previously the Doppler velocity image itself was convolved with
an MDI point-spread-function. Here we make red and blue intensity images from our
Doppler velocity image and a simple limb darkened intensity image, convolve those with
an MDI point-spread-function, and construct the blurred Doppler velocity image from the
difference divided by the sum.

This process yields Doppler velocity images that are visually indistinguishable from
MDI Doppler velocity images. A velocity image comparison is shown in Fig. 1.

\begin{figure}[!ht]
\plotone{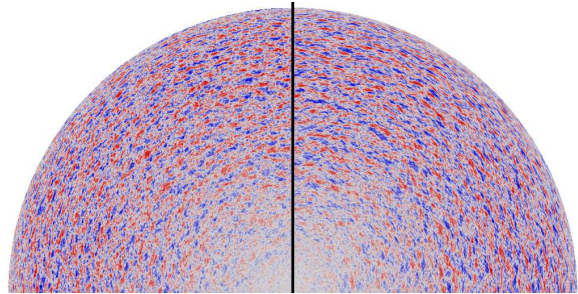}
\caption{
Doppler velocity images (MDI on left, simulation on right) in which the observer's
motion, the convective blue shift, and the Doppler signals due to the axisymmetric
flows --- differential rotation and meridional circulation --- are removed to better
reveal the cellular structures (supergranules). The spatial characteristics are visually indistinguishable.
}
\end{figure}

The velocity pattern is evolved in time by introducing changes to the spectral coefficients
based on two processes - the advection by the axisymmetric flows (differential
rotation and meridional flow) and random processes that lead to the finite lifetimes
of the cells.

The advection is governed by an advection equation

\begin{equation}
{\partial w \over \partial t} = - {V(\theta) \over r} {\partial w \over \partial \theta}
- {U(\theta) \over {r \sin \theta}} {\partial w \over \partial \phi}
\end{equation}

\noindent where $w$ is a velocity component, $U(\theta) = r \sin \theta \Omega(\theta)$
gives the differential rotation profile and $V(\theta)$ gives the meridional
flow velocity profile.
Representing $w$ as a series of spherical harmonic components (Eqs. 1-3) and
projecting this advection equation onto a single spherical harmonic gives a series
of coupled equations for the evolution of the spectral coefficients (Appendix A).
Solid body rotation simply introduces a constantly increasing phase for each coefficient. Differential rotation couples the phase change in one spectral coefficient to spectral
coefficients with wavenumbers $\ell \pm 2$ and $\ell \pm 4$ for differential rotation
of the form

\begin{equation}
\Omega(\theta) = \Omega_0 + \Omega_2 \cos^2 \theta + \Omega_4 \cos^4 \theta
\end{equation}

\noindent while a simple but reasonable meridional flow profile with

\begin{equation}
V(\theta) = V_0 \cos \theta \sin \theta
\end{equation}

\noindent couples one spectral coefficient to spectral coefficients
with wavenumbers $\ell \pm 2$. (Spherical harmonics have fixed latitudinal structure.
Spectral power must pass from one spherical harmonic component to another
in order to move a feature in latitude.)

These cellular flows are embedded in the Sun's surface shear layer.
We approximate the change in the rotation rate in the outermost 5\%
of the Sun as reported by \citet{Howe07} with

\begin{equation}
\Omega(r,\theta) = \Omega(\theta) f(r)
\end{equation}

\noindent where

\begin{equation}
f(r) = 1 + 0.038 \left[ 1 - e^{-55 (1 - r/R_\odot)} \right]
\end{equation}

\noindent and the latitude dependence is given by

\begin{equation}
{\Omega_0 \over 2\pi} = 452 \ \rm nHz \ (14.07 \degr \ day^{-1})
\end{equation}

\begin{equation}
{\Omega_2 \over 2\pi} = -55 \ \rm nHz \ (-1.75 \degr \ day^{-1})
\end{equation}

\begin{equation}
{\Omega_4 \over 2\pi} = -75 \ \rm nHz \ (-2.30 \degr \ day^{-1})
\end{equation}

\noindent Assuming that the cells extend to depths similar to their horizontal dimensions,
and that they are advected at flow rates representative of that depth, Eq. 9 is
transformed into a function of $\ell$ with

\begin{equation}
f(\ell) = 1 + 0.038 \left[ 1 - e^{-90 \pi/\ell} \right]
\end{equation}

\noindent This shear layer profile is illustrated in Fig. 2 along with the gradients
expected from theoretical arguements for flows that conserve angular momentum.
We assume a meridional flow which is constant with depth across this layer and
has a latitude dependence characterized by

\begin{equation}
V_0 = -30 \ \rm m \ s^{-1}
\end{equation}

\noindent which gives a peak meridional flow velocity of $15 \ \rm m \ s^{-1}$ at
$45\degr$ latitude.

\begin{figure}[!ht]
\plotone{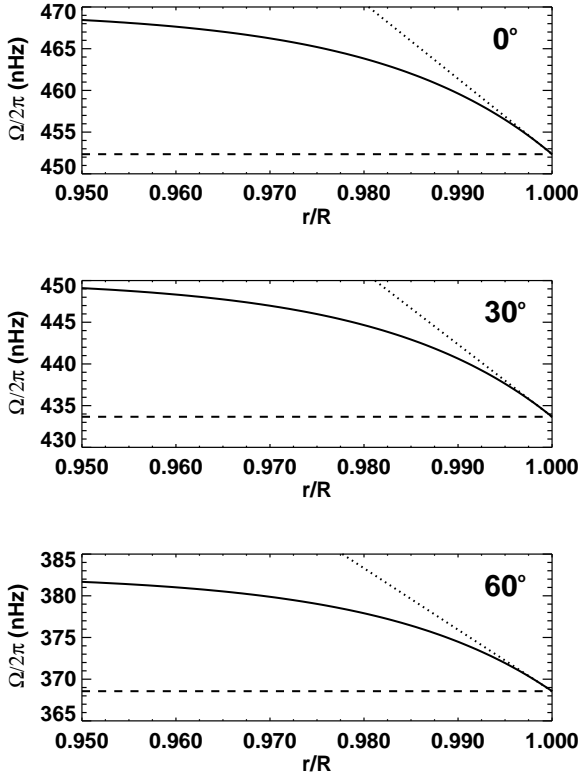}
\caption{
The assumed rotation rate as a function of radius in the surface shear layer for three
different latitudes.
The surface rotation rate at each latitude is indicated by the dashed lines.
The variations in rotation rate for flows that conserve angular
momentum from the surface inward are indicated by the dotted lines.
}
\end{figure}

The finite lifetimes for the cells are simulated by introducing random perturbations
to the spectral coefficient phases. The size of these perturbations
increases with wavenumber to give shorter lifetimes to smaller cells with

\begin{equation}
\delta \Phi_\ell^m \propto \sqrt{\Delta t / \tau(\ell)}
\end{equation}

\noindent where $\delta\Phi_\ell^m$ is the change in phase for a complex spectral
coefficient of degree $\ell$ and order $m$, $\Delta t$ is the time interval
between simulated Doppler images, and $\tau(\ell)$ is proportional to the
lifetime for a spectral component of degree $\ell$. Lifetimes are well
approximated by a turn-over time for turbulent convective flows.
The cellular flow velocities are roughly proportional to $\ell$
while their diameters are inversely proportional to $\ell$.
The turn-over times should then be inversely proportional to $\ell^2$.
We find a reasonable fit to the data using

\begin{equation}
\tau(\ell) = 6.5 {100^2 \over \ell^2} \rm{hrs}
\end{equation}

\section{The Analyses}

Several anaylsis programs were applied to both the MDI data and the simulated data.
Power spectra were obtained to characterize and compare the distribution of cell
sizes and flow velocities. The rotation of the Doppler pattern was determined using
multiple techniques based on previous studies. The meridional flow was measured
based on the movement of the Doppler pattern and cell lifetimes were estimated
from cross-correlation analyses.

\subsection{Convection Spectra} 

Convection spectra for individual images were obtained using the methods described
by \citet{Hathaway87} and \citet{Hathaway92} -- the Doppler signal due to the motion
of the observer is removed, the convective blue shift signal is identified and removed,
the data is mapped to heliographic coordinates, the axisymmetric flow signals due
to differential rotation and meridional circulation are identified and removed,
and the remaining signal is projected onto spherical harmonics. The averaged spectra
from the 60-day 1996 MDI Dynamics Run and from our 10-day simulated data run
are shown in Fig. 3.

The match between these spectra is obtained by adjusting the input
spectrum for the simulated data. This spectrum contains two Lorentzian-like
spectral components -- a supergranule component centered on $\ell \sim 110$ with
a width of about 100 and a granule component centered on $\ell \sim 4000$ with a
width of about 3000. The MDI spectrum is well matched with just these two components
without the addition of the mesogranule component suggested by \citet{November_etal81}.
In fact, we have a distinct {\em dip} in the input spectrum at wavenumbers $\ell \sim 500$
that should be representative of mesogranules. \citet{Hathaway_etal00} showed that
this dip is easily seen in the observed spectra from the MDI high resolution data.

\begin{figure}[!ht]
\plotone{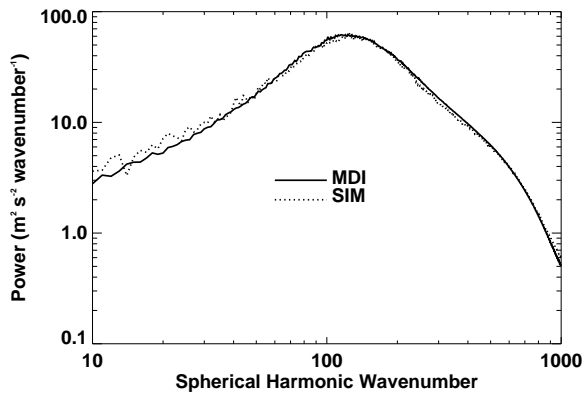}
\caption{
Power spectra from spherical harmonic analyses of MDI data (solid line) and simulated
data (dotted line). The spectral match indicates that both datasets have very similar spatial
structures. (The rapid drop in power for wavenumbers beyond 600 is largely due to instrumental blurring.)
}
\end{figure}

\subsection{Rotation}

Rotation information is obtained when additional analyses are applied to the data
after it has been mapped onto heliographic coordinates.
Longitudinal strips of this data, centered on latitudes from $75^\circ$ south
to $75^\circ$ north, were cross-correlated with corresponding strips
from later images as was done by \citet{Duvall80} and by \citet{SnodgrassUlrich90}.
The shift in the location of the cross-correlation peak divided by the
time difference gives the rotation rate.

Fig. 4 shows the differential rotation from the cross-correlation analysis.
The profiles from the simulated data match those from the MDI data at all but
the highest latitudes.
Both datasets show faster rotation rates for longer time lags as noted by \citet{Duvall80}
and by \citet{SnodgrassUlrich90}. The match between the MDI data and the simulated
data indicates that the latitudinal differential rotation profile is fairly well represented
by Eq. 6 with the coefficients given by Eqns. 10-12 derived from the helioseismic
studies of \citet{Howe07}. (The simulated data would be a better match at high
latitudes for $\Omega_4 = -85$ nHz.) The MDI data does show slightly faster equatorial
rotation. This indicates that either Eq. 13 should be modified to give faster rotation
at the lower wavenumbers or Eq. 16 should be modified to give shorter lifetimes at the
higher wavenumbers.

\begin{figure}[!ht]
\plotone{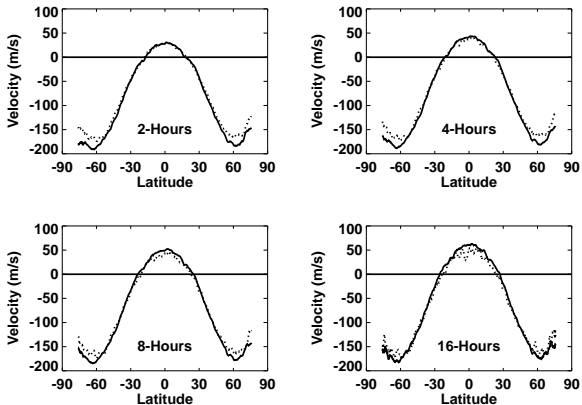}
\caption{
Differential rotation profiles from cross-correlation analyses of MDI data (solid lines)
and simulated data (dotted lines) for four different time lags between Doppler images --
2, 4, 8, and 16 hours. All profiles are well matched at all time lags and
at all but the highest latitudes.
}
\end{figure}

We reproduced the analysis of \citet{BeckSchou00}.
Lines of data from the heliographic maps were Fourier analyzed in longitude
and those spectral coefficents were then Fourier anaylzed in time over
10-day intervals. The power spectra were averaged over all latitude lines
between $\pm 9\degr$. The rotation rate for each wavenumber $\ell$ was determined
by finding the temporal frequency of the center of gravity of the power using
a frequency window of $\approx 58 \mu \rm Hz$ (and iterating on the position
of that window four times) and then dividing the temporal frequency by the wavenumber.

Fig. 5 shows the equatorial rotation rates as functions of wavenumber for both
the MDI data and the simulated data. Here again we find a good match for all but
the lowest wavenumbers. This indicates that Eq. 13 gives a good wavenumber dependence
for the rotation rate. The drop in rotation rate for the MDI data at wavenumbers
below ~30 is due to imaging artifacts that introduce signal at low temporal
frequencies which can overlap with the rotation signal at low spatial frequencies.
The effects of the line-of-sight projection on measured rotation rates, as discussed
by \citet{Hathaway_etal06}, are evident in the increase in the observed rotation rates
above the input rates from Eq. 13 for wavenumbers below $\sim 100$.
Although the higher noise level from the single 10-day simulation makes precise
comparisons difficult, it does appear that the MDI data might be better fit if
the rotation rates given by Eq. 13 were somewhat higher at wavenumbers below $\sim 100$.

\begin{figure}[!ht]
\plotone{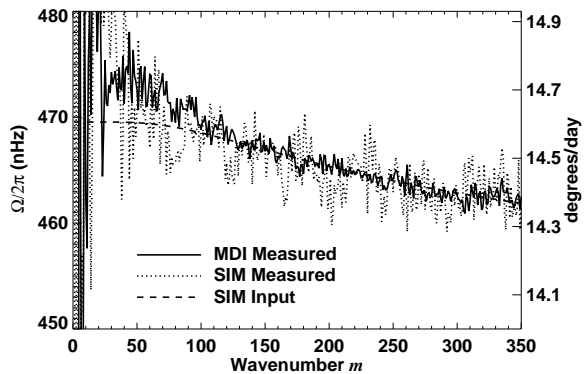}
\caption{
Equatorial rotation rates as functions of wavenumber from analyses of 6 10-day sets of
MDI data (solid line) and a single 10-day set of simulated data (dotted line). The
equatorial rotation rates used as input (Eq. 13) are shown with the dashed line.
The rotation rates match very well at all but the lowest wavenumbers.
Rotation rates faster than the input rates at wavenumbers below about 100 are due
to projection effects.
}
\end{figure}

We have also reproduced one of the analyses of \citet{Schou03}.
Lines of data from the heliographic maps are apodized and then multiplied by
longitude dependent weighting functions designed to remove the Doppler
projection effect and to isolate either longitudinal motions or latitudinal motions.
The weighting used to isolate longitudinal flows is based on the final term in
Eq. 4 but with $B_0 = 0$ and a constant of 0.01 added to $\sin \phi$ to avoid
division by zero at the central meridian.
The lines are shifted in longitude according to the differential rotation rate and then
Fourier analysed in space and time to obtain ``$k-\omega$'' diagrams.
\citet{Schou03} noted that these diagrams show power for both prograde
($\omega/k < 0$) and retrograde ($\omega/k > 0$) motion but with
excess power in the prograde components.

Fig. 6 shows the $k-\omega$ diagrams from 10 days of the MDI data and from
the simulated data for latitudes between $\pm4.5\degr$.
Both show prograde and retrograde components as similarly
shaped wedges of enhanced power extending from the origin to both positive
and negative longitudinal wavenumbers. However, the MDI data does show excess
power for prograde motion (power excess extending from the upper left to the lower right)
that is not evident in the simulated data. (Note that
running this analysis on the full 60-day run of MDI data shows a clear
separation between prograde and retrograde motions.)

\begin{figure}[!ht]
\plotone{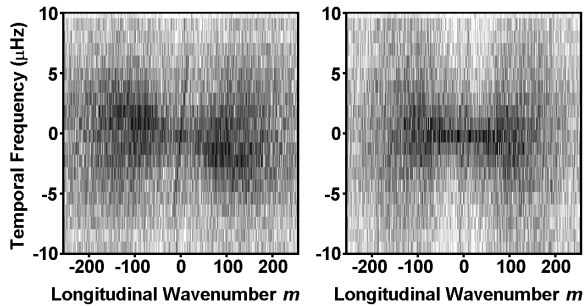}
\caption{
The $k-\omega$ diagrams from analyses of MDI data (left) and simulated data (right).
Darker shades indicate more power. Both datasets show prograde and retrograde
components with similar power and limits but the MDI data shows excess power
in the prograde components (power in upper left and lower right quadrants).
The thin line from upper right to lower left in the MDI data is produced by
imaging artifacts which appear to move retrograde at the Sun's rotation rate.
}
\end{figure}

\subsection{Meridional Flow}

The meridional motion of the Doppler pattern can be measured by the cross-correlation
method used to find the differential rotation profiles illustrated in Fig. 4 or by using
the apodizing/weighting method described by \citet{Schou03}. Fig. 7 shows a comparison
of the meridional flow profiles obtained from the MDI data and the simulated data
when we take $V_0 = -30 \ \rm m\ s^{-1}$, representing a poleward meridional flow with peak velocity of $15 \ \rm m\ s^{-1}$. The meridional flow we find for the Doppler features in
the simulation very closely matches the input meridional flow (dashed lines in Fig. 7).
This indicates that the projection effects that produce the super-rotation of the
Doppler features do not impact the measurements of the meridional flow.

\begin{figure}[!ht]
\plotone{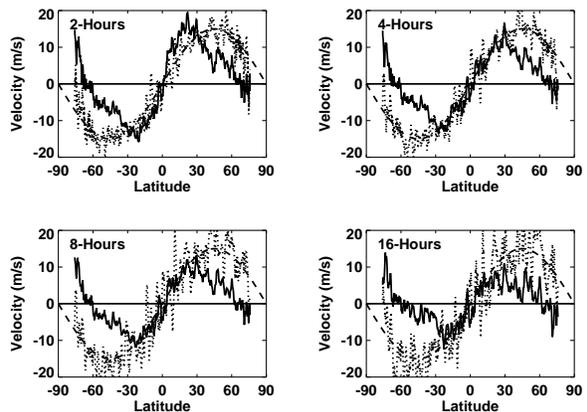}
\caption{
Meridional flow profiles from cross-correlation analyses of MDI data (solid lines) and
simulated data (dotted lines) for four different time lags between Doppler images --
2, 4, 8, and 16 hours. The meridional flow profile input to the simulation is
shown by the dashed lines.
}
\end{figure}

It is clear that the actual meridional flow of the supergranules is more complicated
than that used in our simulation. The meridional flow profiles for the supergranules
shown in Fig. 7 for the MDI data agree very well with those obtained by \citet{Gizon_etal03}
and by \citet{Schou03} using somewhat different methods. The MDI data indicates the
presence of counter-cells in the polar regions during the 1996 data run that would be
better represented by

\begin{equation}
V(\theta) = (50 \cos \theta - 60 \cos^3 \theta) \sin \theta
\end{equation}

\noindent In addition the flow appears to weaken for longer time differences between
the cross-correlated data. This suggests the the meridional flow in the surface shear
layer decreases in amplitude with depth.

\subsection{Lifetimes}

The lifetimes of the cells can be estimated by comparing the strength of the
cross-correlation coefficients for the Doppler features as functions of the time
interval between Doppler images.
Fig. 8 shows the strength of the correlations as functions of latitude and time lag
for both the MDI data and the simulated data.
The profiles are similar in shape but slightly flatter with the MDI data.
The correlation coefficients are well matched at the equator for time differences
of 4-hours and 16-hours but the simulation shows stronger correlations at 2-hours
and weaker correlations at 8-hours. 
This suggests that the formulae (Eqns. 15 and 16) for the random phase changes
to the spectral coefficients needs slight modification. In particular the cellular
features that dominate at $\Delta t = 2$ hours need shorter lifetimes.

\begin{figure}[!ht]
\plotone{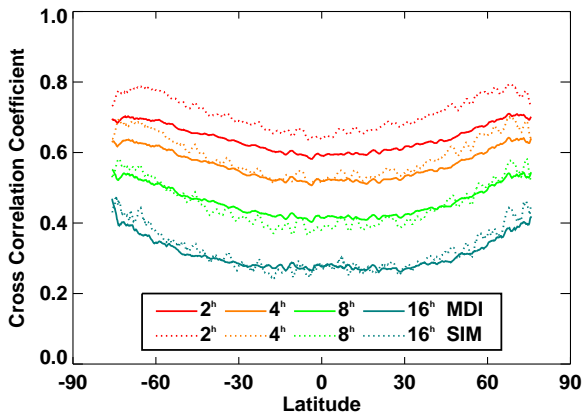}
\caption{
Cross-correlation peak profiles from analyses of MDI data (solid lines) and simulated
data (dotted lines). The profiles are well matched at $\Delta t = 4$  and
$\Delta t = 16$ hours but with slight differences at $\Delta t = 2$  and
$\Delta t = 8$ hours.
}
\end{figure}

\section{Conclusions}

We have produced simulated data in which the cellular structures (supergranules) are advected
by differential rotation and meridional flow and evolved by uncorrelated random changes.
When we compare results from analyses of these data with those from the MDI
data we find that the simulated data exhibit much of the same characteristics as the MDI
data -- the visual structures, the power spectra, the rotation, the meridional flow,
and the evolution rates all match fairly closely.
While some of the rotation characteristics have been attributed to wave-like properties
by \citet{BeckSchou00}, \citet{Gizon_etal03}, and \citet{Schou03}, our simulated data is simply
advected by a zonal flow (differential rotation) with speeds that never exceed those determined from helioseismology as reported by \citet{Howe07}.

While the similarities between the simulated data and the MDI data are strong,
the differences are important and interesting. The rotation variations with depth
and latitude were chosen to closely match those shown in Fig 1. of \citet{Howe07}.
The differences in rotation rate seen at high latitudes in our Fig. 4 suggest
that the magnitude of $\Omega_4$ should be increased from -75 nHz to -85 nHz.
The differences in rotation rate seen at low wavenumbers in our Fig. 5 suggest
an increase in rotation across the surface shear layer of about 2 nHz at
the equator.

The lack of excess power in the prograde components represented
in Fig 6 for the simulation and the lack of evidence for a split between the
prograde and retrograde components indicate differences in the evolution
of the pattern. The manner in which we evolve the cells in the simulation
is not fully satisfactory. The phase (and amplitude)
changes associated with the evolving cells should replicate their advection by
larger (nonaxysymmetric) flows \citep{WilliamsCuntz09},
their break-up into smaller cells, and their
joining to form larger cells. The random changes in phase that we introduce
to evolve the cells produces random offsets that are larger for larger
wavenumbers and, when modulated by the power spectrum shown in Fig 3,
produces the wedge of power shown in Fig 6 for the simulation.
The split between prograde and retrograde movement seen in the MDI data suggests
that the motions are not random but preferentially east-west.
\citet{Lisle_etal04} found that supergranules appear to have a persistent
north-south alignment and suggested that this was due to a north-south
elongation of larger, giant cells that advect the supergranules to their
boundaries. The rotational constraints on these giant cells 
[cf. \citet{Miesch_etal08} and references therein] should in-fact
give a preference for east-west flows near the equator and may produce
an added preference for prograde flows. Note that this interpretation
of the results does not involve wave-like properties for supergranules --
but instead invokes simple advection by larger flows.

The advection of the supergranules by the meridional flow shows considerable
promise. The fact that we fully and accurately recover the input velocity
profile indicates that the latitudinal movement of supergranules provides
new information on the meridional flow. The results with the MDI data indicate
the presence of counter-cells (equatorward flow) at latitudes above about
$65\degr - 70\degr$.
These results also indicate a meridional flow speed that matches that found for the
small magnetic elements as measured by \citet{Komm_etal93} and by \citet{HathawayRightmire10}
but is significantly slower than that reported in the helioseismic determination
of \citet{Giles_etal97} and \citet{Basu_etal99}. The indication of a decrease in meridional flow speed with depth is also at odds with those helioseismic results
but in agreement with the simulations of \citet{Hathaway82}.

Magnetic element positions are known to coincide with the boundaries of
supergranules and move as the supergranule boundaries evolve. This was shown
fairly explicitly by \citet{Lisle_etal00}.
This suggests that the advection of the supergranules by the axisymmetric flows
should be directly related to the advection of the magnetic elements.
The differential rotation itself (relative to the Carrington rotation) has a velocity
range from $\sim 40 \ \rm m\ s^{-1}$ prograde to $\sim 180 \ \rm m\ s^{-1}$ retrograde
while the meridional flow has a peak velocity of only $\sim 11 \ \rm m\ s^{-1}$.
These weak flow velocities are no match for the flows of several 100 $\rm m\ s^{-1}$
in the supergranules. The magnetic elements should be quickly carried to the boundaries and
ultimately the interstices of the supergranules where the direct action of the
differential rotation and meridional flow are too weak to dislodge them.
The magnetic elements should experience differential rotation and meridional flow only
to the extent that the supergranules themselves are advected by these flows
at velocities representitive of deeper layers.

The variation in the differential rotation and meridional flow with depth and latitude
within the surface shear layer needs further examination. We have already argued for some
changes from what was assumed in our simulation. In addition, we note that our association
of different depths with different wavenumbers $\ell$ as indicated in Eqns. 9 and 13
suggests that cells with diameters $D = 2 \pi R_\odot/ \ell$ are advected by flows
at depth $R_\odot - r \approx 0.82 D$. This would make typical supergranules extend
to depths of 30 Mm. While this is deeper than suggested by the helioseismic
investigations of \citet{Duvall_etal97} and \citet{Woodard07}, those investigations
were limited to layers not much deeper than 7 Mm. Shallower cells would suggest
that the surface shear layer is shallower as well.

\acknowledgements
We would like to thank: NASA for its support of this research through a grant
from the Heliophysics Guest Investigator Program to NASA Marshall Space Flight Center;
SOHO, which is a project of international cooperation between ESA and NASA; and
John Beck, who prepared the 31-minute filtered SOHO/MDI data used in this paper.

\appendix

\section{Spectral Coefficient Changes Due to Advection}

The changes in the spectral coefficients produced in the advection of the pattern
by the axisymmetric flows are best illustrated using the radial component of the flow pattern as given by

\begin{equation}
w(\theta,\phi,t) = \sum_{\ell=0}^{lmax} \sum_{m=-\ell}^\ell R_\ell^m(t) P_\ell^m(x) e^{i m \phi}
\end{equation}

\noindent where the time-varying spectral coefficient is $R_\ell^m(t)$, $x=\cos \theta$, and $P_\ell^m$
is an Associated Legendre polynomial of angular degree $\ell$ and azimuthal order $m$.
The Associated Legendre polynomials are normalized such that

\begin{equation}
\int_{-1}^1 P_\ell^m(x) P_{\ell^\prime}^{m^\prime}(x) dx = \delta_\ell^{\ell^\prime}
\delta_m^{m^\prime}
\end{equation}

The advection of this flow pattern by the axisymmetric flows is given by

\begin{equation}
{{\partial w} \over {\partial t}} = - {V \over r} {{\partial w} \over {\partial \theta}}
- {U \over {r \sin \theta}} {{\partial w} \over {\partial \phi}} 
\end{equation}

\noindent The spherical harmonic representation gives explicit expressions for the spatial
derivatives with

\begin{equation}
{{\partial w} \over {\partial \theta}} = \sum_{\ell=0}^{lmax} \sum_{m=-\ell}^\ell R_\ell^m(t)
{1 \over \sin \theta} \left[{\ell \over A_{\ell + 1}^m} P_{\ell + 1}^m(x) - 
{(\ell + 1) \over A_\ell^m} P_{\ell - 1}^m(x)\right]
e^{i m \phi}
\end{equation}

\noindent and

\begin{equation}
{{\partial w} \over {\partial \phi}} = \sum_{\ell=0}^{lmax} \sum_{m=-\ell}^\ell R_\ell^m(t)
i m P_\ell^m(x) e^{i m \phi}
\end{equation}

\noindent where

\begin{equation}
A_\ell^m = \left[{{(2\ell + 1) (2\ell - 1)} \over
{(\ell + m) (\ell - m)}}\right]^{1 \over 2}
\end{equation}

\noindent Taking the meridional velocity as

\begin{equation}
V(\theta) = V_0 \cos \theta \sin \theta
\end{equation}

\noindent and the azimuthal velocity as

\begin{equation}
U(\theta) = r \sin \theta (\Omega_0 + \Omega_2 \cos^2 \theta + \Omega_4 \cos^4 \theta)
\end{equation}

\noindent and then projecting eq. A3 onto $Y_{\ell^\prime}^{m^\prime}$ gives

\begin{eqnarray}
{{\partial R_{\ell^\prime}^{m^\prime}} \over {\partial t}} & = & - {V_0 \over r} \sum_{\ell=m^\prime}^{lmax} R_\ell^{m^\prime} 
\int_{-1}^1 \left[{\ell \over A_{\ell + 1}^{m^\prime}} x P_{\ell + 1}^{m^\prime}(x) P_{\ell^\prime}^{m^\prime}(x) -
{(\ell + 1) \over A_\ell^{m^\prime}} x P_{\ell - 1}^{m^\prime}(x) P_{\ell^\prime}^{m^\prime}(x)\right]dx \nonumber \\
& & -i m^\prime \Omega_0 R_{\ell^\prime}^{m^\prime}
-i m^\prime \Omega_2 \sum_{\ell=m^\prime}^{lmax} R_\ell^{m^\prime}  \int_{-1}^1 x^2 P_\ell^{m^\prime}(x) P_{\ell^\prime}^{m^\prime}(x) dx \nonumber \\
& & -i m^\prime \Omega_4 \sum_{\ell=m^\prime}^{lmax} R_\ell^{m^\prime} \int_{-1}^1 x^4 P_\ell^{m^\prime}(x) P_{\ell^\prime}^{m^\prime}(x) dx
\end{eqnarray}

\noindent The products of $x$, $x^2$, and $x^4$ with the Associated Legendre polynomials
can be replaced with expressions containing only Associated Legendre polynomials of the
same azimuthal order, $m$, using the recursion relation

\begin{equation}
P_\ell^m(x) = A_\ell^m x P_{\ell - 1}^m(x) - B_\ell^m P_{\ell - 2}^m(x)
\end{equation}

\noindent where $A_\ell^m$ is given by eq. A6 and

\begin{equation}
B_\ell^m = \left[{{(2\ell + 1) (\ell + m - 1) (\ell - m - 1)} \over
{(2\ell - 3) (\ell + m) (\ell - m)}}\right]^{1 \over 2}
\end{equation}

\noindent Eq. A9 then reduces to (after dropping the primes)

\begin{eqnarray}
{{\partial R_\ell^m} \over {\partial t}} & = & - {V_0 \over r} \left[
{{(\ell - 2)} \over {A_\ell^m} A_{\ell - 1}^m} R_{\ell - 2}^m
+ \left[ {{\ell B_{\ell+2}^m} \over {A_{\ell+2}^m A_{\ell+1}^m}}
- {{(\ell+1)} \over {A_\ell^m A_\ell^m}} \right] R_\ell^m 
- {{(\ell + 3) B_{\ell + 2}^m} \over {A_{\ell + 2}^m A_{\ell + 2}^m}} R_{\ell + 2}^m 
\right] - i m \Omega_0 R_\ell^m
\nonumber \\
& & - i m \Omega_2 \left[ { 1 \over {A_\ell^m A_{\ell-1}^m}} R_{\ell-2}^m
+ \left[ {B_{\ell+2}^m \over {A_{\ell+2}^m A_{\ell+1}^m}} 
+ {B_{\ell+1}^m \over {A_{\ell+1}^m A_\ell^m}} \right] R_\ell^m
+ {B_{\ell+3}^m \over {A_{\ell+3}^m A_{\ell+2}^m}} B_{\ell+2}^m R_{\ell+2}^m
\right] \nonumber \\
& & - i m \Omega_4 \left[ {1 \over {A_\ell^m A_{\ell-1}^m}}
{1 \over {A_{\ell-2}^m A_{\ell-3}^m}} \right] R_{\ell-4}^m
\nonumber \\
& & - i m \Omega_4 \left[ {B_{\ell+2}^m \over {A_{\ell+2}^m A_{\ell+1}^m}}
+ {B_{\ell+1}^m \over {A_{\ell+1}^m A_{\ell}^m}}
+ {B_{\ell}^m \over {A_{\ell}^m A_{\ell-1}^m}}
+ {B_{\ell-1}^m \over {A_{\ell-1}^m A_{\ell-2}^m}}
\right] {1 \over {A_{\ell}^m A_{\ell-1}^m}} R_{\ell-2}^m
\nonumber \\
& & - i m \Omega_4 \left[
{B_{\ell+3}^m \over {A_{\ell+3}^m A_{\ell+2}^m}} 
+ {B_{\ell+2}^m \over {A_{\ell+2}^m A_{\ell+1}^m}} 
+ {B_{\ell+1}^m \over {A_{\ell+1}^m A_{\ell}^m}}
\right] {B_{\ell+2}^m \over {A_{\ell+2}^m A_{\ell+1}^m}} R_\ell^m
\nonumber \\
& & - i m \Omega_4 \left[
{B_{\ell+2}^m \over {A_{\ell+2}^m A_{\ell+1}^m}} 
+  {B_{\ell+1}^m \over {A_{\ell+1}^m A_{\ell}^m}}
+  {B_{\ell}^m \over {A_{\ell}^m A_{\ell-1}^m}}
\right] {B_{\ell+1}^m \over {A_{\ell+1}^m A_{\ell}^m}} R_\ell^m
\nonumber \\
& & - i m \Omega_4 \left[ {B_{\ell+4}^m \over {A_{\ell+4}^m A_{\ell+3}^m}}
+ {B_{\ell+3}^m \over {A_{\ell+3}^m A_{\ell+2}^m}}
+ {B_{\ell+2}^m \over {A_{\ell+2}^m A_{\ell+1}^m}}
+ {B_{\ell+1}^m \over {A_{\ell+1}^m A_{\ell}^m}}
\right] {B_{\ell+3}^m \over {A_{\ell+3}^m A_{\ell+2}^m}} B_{\ell + 2}^m R_{\ell+2}^m
\nonumber \\
& & - i m \Omega_4 \left[ + {B_{\ell+5}^m \over {A_{\ell+5}^m A_{\ell+4}^m}}
 {B_{\ell+3}^m \over {A_{\ell+3}^m A_{\ell+2}^m}}
B_{\ell+4}^m B_{\ell+2}^m \right] R_{\ell+4}^m
\end{eqnarray}

\noindent A common factor in Eq. A12 is of the form

\begin{equation}
{B_\ell^m \over {A_\ell^m A_{\ell-1}^m}} = {{(\ell + m - 1)(\ell - m - 1)}
\over {(2 \ell - 1)(2 \ell - 3)}}
\end{equation}

\noindent The dominant terms on the RHS of Eq. A12 are those for the direct rotation
of the $R_\ell^m$ coefficients. These terms are best handled analytically by taking

\begin{equation}
R_\ell^m = \mathcal{R}_\ell^m e^{i m \Omega_\ell^m t}
\end{equation}

\noindent where
\begin{eqnarray}
\Omega_\ell^m & = &\Omega_0 +
\Omega_2 \left[ {B_{\ell+2}^m \over {A_{\ell+2}^m A_{\ell+1}^m}} 
+ {B_{\ell+1}^m \over {A_{\ell+1}^m A_\ell^m}} \right] + \nonumber \\
& & \Omega_4 \left[
{B_{\ell+3}^m \over {A_{\ell+3}^m A_{\ell+2}^m}} 
+ {B_{\ell+2}^m \over {A_{\ell+2}^m A_{\ell+1}^m}} 
+ {B_{\ell+1}^m \over {A_{\ell+1}^m A_{\ell}^m}}
\right] {B_{\ell+2}^m \over {A_{\ell+2}^m A_{\ell+1}^m}} + \nonumber \\
& & \Omega_4 \left[
{B_{\ell+2}^m \over {A_{\ell+2}^m A_{\ell+1}^m}} 
+  {B_{\ell+1}^m \over {A_{\ell+1}^m A_{\ell}^m}}
+  {B_{\ell}^m \over {A_{\ell}^m A_{\ell-1}^m}}
\right] {B_{\ell+1}^m \over {A_{\ell+1}^m A_{\ell}^m}}
\end{eqnarray}

\noindent With this substitution, Eq. A12 becomes an equation for the evolution of
$\mathcal{R}_\ell^m$ in which the meridional flow (terms multiplied by $V_0$)
couples the amplitude of the $\ell$ component to the $\ell \pm 2$ components
while the differential rotation (terms multiplied by $\Omega_2$ and $\Omega_4$)
couples the phase of the $\ell$ component to the $\ell \pm 2$ and $\ell \pm 4$
components.
These spectral coefficients can be evolved in time using Eq. A12 with a fourth-order
Runge-Kutta scheme. The solutions are stable and well behaved for quarter-hour
time steps over at least 10-days for $\ell$ upto at least 1500.

\end{document}